\documentclass{article}

\usepackage{microtype}
\usepackage{graphicx}
\usepackage{subfigure}
\usepackage{booktabs} 
\usepackage{hyperref}

\usepackage[accepted]{icml2025}

\usepackage{amsmath}
\usepackage{amssymb}
\usepackage{mathtools}
\usepackage{amsthm}
\usepackage[most]{tcolorbox}
\usepackage{xcolor}
\usepackage{colortbl}
\usepackage{threeparttable}

\usepackage[capitalize,noabbrev]{cleveref}

\theoremstyle{plain}

\theoremstyle{definition}

\theoremstyle{remark}

\icmltitlerunning{CosmoPaperQA: Benchmarking RAG Agents for Astronomical Literature}

\begin{document}

\twocolumn[
\icmltitle{
Evaluating Retrieval-Augmented Generation Agents for Autonomous Scientific Discovery in Astrophysics}


\icmlsetsymbol{equal}{*}

\begin{icmlauthorlist}
\icmlauthor{Xueqing Xu}{equal,camphy}
\icmlauthor{Boris Bolliet}{equal,camphy,kavli}
\icmlauthor{Adrian Dimitrov}{equal,camphy}
\icmlauthor{Andrew Laverick}{camphy}
\icmlauthor{Francisco Villaescusa-Navarro}{flatiron,princ}
\icmlauthor{Licong Xu}{kavli,ia}
\icmlauthor{Íñigo Zubeldia}{kavli,ia}
\end{icmlauthorlist}

\icmlaffiliation{ia}{Institute of Astronomy, University of Cambridge, Cambridge, United Kingdom}
\icmlaffiliation{kavli}{Kavli Institute for Cosmology, University of Cambridge, Cambridge, United Kingdom}

\icmlaffiliation{camphy}{Department of Physics, University of Cambridge, Cambridge, United Kingdom}

\icmlaffiliation{flatiron}{Center for Computational Astrophysics, Flatiron Institute, New York, NY, USA}
\icmlaffiliation{princ}{Department of Astrophysical Sciences, Princeton University, Princeton, NJ, USA}

\icmlcorrespondingauthor{Xueqing Xu}{april.xuq@gmail.com}
\icmlcorrespondingauthor{Boris Bolliet}{bb667@cam.ac.uk}

\icmlkeywords{Machine Learning, Deep Learning, Neural Networks, ICML}

\vskip 0.3in
]

\printAffiliationsAndNotice{\icmlEqualContribution} 

\begin{abstract}
We evaluate 9 Retrieval Augmented Generation (RAG) agent configurations on 105 Cosmology Question-Answer (QA) pairs that we built specifically for this purpose.\footnote{\href{https://huggingface.co/datasets/ASTROANTS/CosmoPaperQA}{https://huggingface.co/datasets/ASTROANTS/CosmoPaperQA}} The RAG configurations are manually evaluated by a human expert, that is, a total of 945 generated answers were assessed. We find that currently the best RAG agent configuration is with OpenAI embedding and generative model, yielding 91.4\% accuracy. Using our human evaluation results we calibrate LLM-as-a-Judge (LLMaaJ) system which can be used as a robust proxy for human evaluation. These results allow us to systematically select the best RAG agent configuration for multi-agent system for autonomous scientific discovery in astrophysics (e.g., \texttt{cmbagent}\footnote{\href{https://github.com/CMBAgents/cmbagent}{https://github.com/CMBAgents/cmbagent}} presented in a companion paper) and provide us with an LLMaaJ system that can be scaled to thousands of cosmology QA pairs. We make our QA dataset, human evaluation results, RAG pipelines, and LLMaaJ system publicly available for further use by the astrophysics community.\footnote{\href{https://github.com/CMBAgents/scirag}{https://github.com/CMBAgents/scirag}} 
\end{abstract}

\section{Introduction}
\label{sec:introduction}
The rapid advancements of Large Language Models (LLMs) \cite{liu2024deepseek,bai2025qwen25vltechnicalreport} have opened a new era in automated scientific discovery, where AI systems can conduct independent research and generate scientific insights \cite{lu2024aiscientistfullyautomated}. In cosmology, automated discovery systems are required to synthesize knowledge across collections of scientific literature, computational models, and observational datasets. The successful implementations require AI infrastructure capable of interacting with the knowledge ecosystem utilized by domain experts, and a specialized computational framework that constitutes the methodological foundation. In this work, we focus on the knowledge integration aspect of automated scientific discovery, specifically targeting the information overload in modern astronomy.

While LLMs have demonstrated impressive capabilities in scientific text analysis \cite{zhang2024comprehensivesurveyscientificlarge}, their deployment in critical research scenarios remains constricted \cite{fouesneau2024rolelargelanguagemodels}, by hallucination \cite{Huang_2025} and knowledge cut-off \cite{cheng2024dateddatatracingknowledge}. Retrieval-Augmented Generation (RAG) has emerged as a powerful tool to enhance LLMs' performance with external knowledge \cite{lewis2021retrievalaugmentedgenerationknowledgeintensivenlp} to meet scientific accuracy standards. The efficacy of this approach has been demonstrated in biology, where PaperQA2 RAG Agents \cite{lála2023paperqaretrievalaugmentedgenerativeagent,skarlinski2024languageagentsachievesuperhuman} achieve superhuman performance on LitQA2 \cite{futurehouse2024litqa2}, a benchmark designed to evaluate knowledge synthesis in real research scenarios.

Despite these successes in biology, systematic evaluation of RAG agents in astronomy remains limited by the lack of standardized benchmarks. As annotated by Bowman et al. \cite{bowman}, developing human-annotated benchmarks for doctoral-level scientific research domains remains economically prohibitive. Consequently, evaluation of RAG agents in astronomy is constrained by the absence of authentic evaluation datasets that capture the complexity of real research scenarios.

To address these challenges, we introduce CosmoPaperQA, a high-quality benchmark dataset including 105 expert-curated question-answer pairs derived from five highly-cited cosmological literature. Unlike synthetic benchmarks, CosmoPaperQA captures authentic research scenarios by extracting questions directly from research papers. 

To facilitate a comprehensive and reproducible evaluation of CosmoPaperQA, we develop SciRag, a modular framework designed for systematic integration and benchmarking of multiple RAG Agents for scientific discovery. Our implementation enables evaluation across commercial APIs (OpenAI Assistant, VertexAI Assistant), hybrid architectures (ChromaDB with several embedding models), specialized academic tools (PaperQA2), and search-enhanced systems (Perplexity), providing empirical guidance for optimal RAG configuration selection in scientific contexts.

Our systematic evaluation across SciRag implementations reveals significant performance differences across four configuration categories, with commercial solutions (OpenAI Assistant: 89.5-91.4\%, VertexAI Assistant: 86.7\%) achieving the highest accuracy on CosmoPaperQA. Hybrid architectures (HybridOAIGem: 85.7\%, HybridGemGem: 84.8\%) show competitive performance while significantly reducing operational costs. Academic tools PaperQA2 (81.9\%) show solid performance but lag behind commercial and hybrid SciRag Agents, while baseline approaches (Gemini Assistant: 16.2\%, Perplexity Assistant: 17.1\%) prove insufficient for expert-level scientific inquiry. 

We present four primary contributions that collectively advance the state of RAG evaluation in cosmology:

\textbf{Benchmark Development:} We introduce CosmoPaperQA, a comprehensive benchmark dataset containing 105 expert-validated question-answer pairs. 

\textbf{Implementation Pipeline:} We develop SciRag, a modular framework that enables systematic deployment and reproducible comparison of diverse RAG solutions.

\textbf{Multi-System RAG Performance Analysis:} We conduct a systematic evaluation of nine distinct RAG implementations utilizing high-performing LLMs and embedding models, revealing significant performance variations across different system architectures and cost-efficiency trade-offs for scientific applications.

\textbf{Calibrated AI Judge Evaluation:} We introduce a LLM-as-a-Judge (LLMaaJ) system that matches human expert assessment in astronomy, enabling scalable performance evaluation while maintaining the quality standards required for scientific applications.

\begin{figure*}[!htbp]
    \centering
    \includegraphics[width=1.0\linewidth]{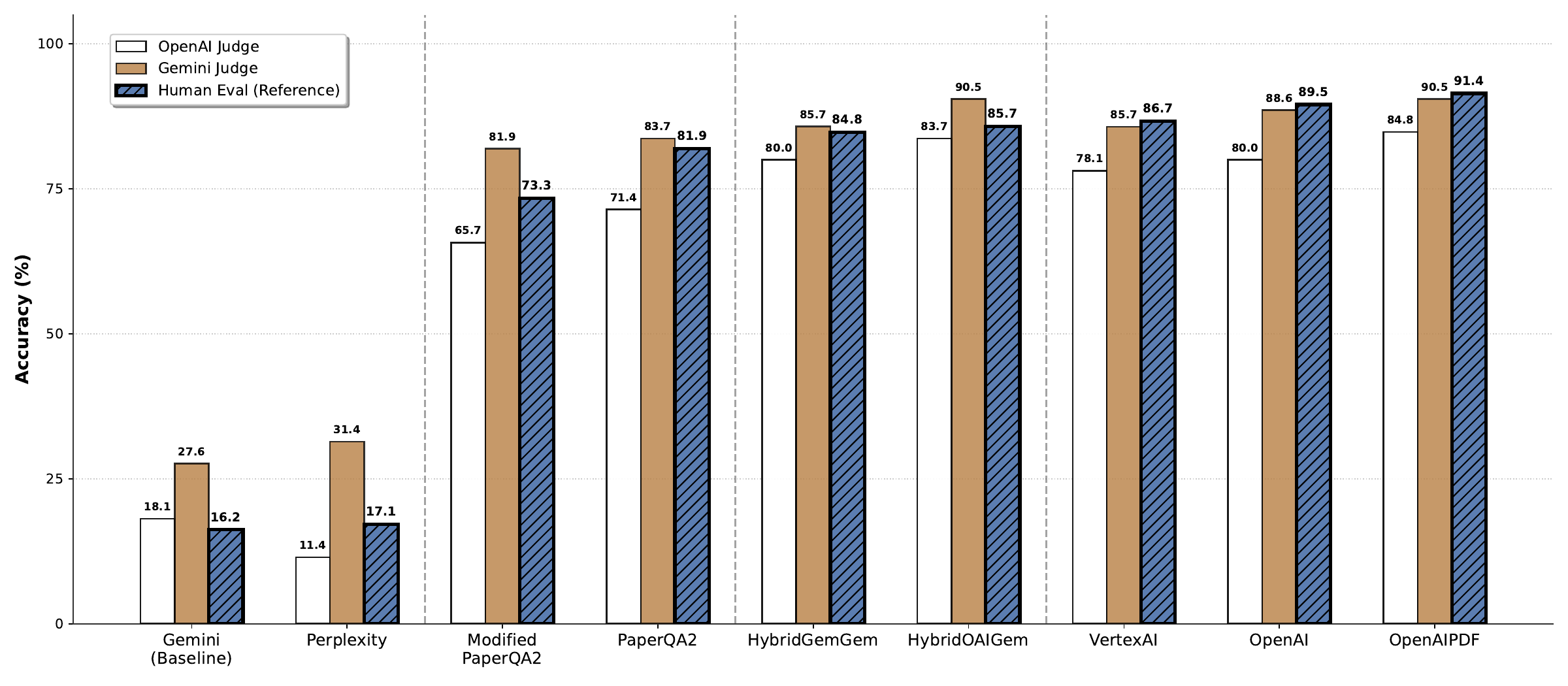}
     \vspace{-20pt} 
    \caption{Performance comparison of SciRag Agents across three evaluation methods. Vertical dashed lines separate different configuration categories: baseline systems (Gemini, Perplexity), academic RAG tools (Modified PaperQA2, PaperQA2), hybrid architectures (HybridGemGem, HybridOAIGem, VertexAI), and commercial solutions (OpenAI, OpenAIPDF). The first two entries (Gemini Baseline and Perplexity) do not perform RAG but simply rely on pre-trained LLM knowledge and, for Perplexity, built-in retrieval tools.}
\label{fig:ai_performance_comparison}
\end{figure*}

\section{Related Work}
\label{sec:related}

\subsection{RAG Agents in Cosmology}
 Recent work has demonstrated the significant potential of LLMs in astronomical research contexts. Ciucă et al. \citep{ciucă2023harnessingpoweradversarialprompting} showed that through in-context learning and adversarial prompting, LLMs can synthesize diverse astronomical information into coherent and innovative hypotheses, while Shao et al. \cite{shao2024astronomicalknowledgeentityextraction} demonstrated their effectiveness in extracting specialized knowledge entities from astrophysics journals using carefully designed prompting strategies. These capabilities have motivated the development of specialized RAG frameworks for astronomy, such as the pathfinder system by Iyer et al. \cite{Iyer_2024}, which implements query expansion, reranking, and domain-specific weighting schemes to enhance retrieval performance in scientific applications.
 
However, the growing deployment of RAG systems in astronomy has highlighted the critical need for systematic evaluation methodologies.  Wu et al. \cite{wu2024designingevaluationframeworklarge} addressed this challenge by proposing a dynamic evaluation framework using a Slack-based chatbot that retrieves information from arXiv astro-ph papers, emphasizing the importance of real-world user interactions over static benchmarks. While their approach provides valuable insights into user behavior and system usability, it relies on user feedback and reaction data rather than systematic performance assessment against validated ground-truth, highlighting a complementary need for standardized benchamrks that can provide consistent, reproducible evaluation metrics across different RAG implementations.

\subsection{Benchmarks and Evaluation in Cosmology}
Existing evaluation falls into two categories, each with some limitations: 

\textbf{Astronomy-Specific Knowledge Benchmarks:} AstroMLab 1 \cite{panastroLab} provides the first comprehensive astronomy-specific evaluation with 4425 AI-generated multiple-choice questions from Annual Review articles. While demonstrating significant performance variations between models with specialized astronomical knowledge, its multiple-choice format and automated question generation limit evaluation to content mastery rather than scientific inquiry workflows. Similarly, Astro-QA \cite{li2025astronomical} provides a structured evaluation with 3082 questions spanning diverse astronomical topics, demonstrating the application of LLMaaJ evaluation in astronomical contexts. However, its synthetic questions limit its ability to assess the complex, open-ended reasoning required for an authentic scientific research workflow.

\textbf{General Scientific Evaluation:}
Broader scientific benchmarks like LitQA2 \cite{futurehouse2024litqa2}, ChemRAG-Toolkit \cite{chembenchmark}, ScisummNet \cite{yasunaga2019scisummnetlargeannotatedcorpus} are designed for other scientific domains and may not capture astronomy-specific challenges such as mathematical reasoning about cosmological models, and interpretation of observational constraints.

\section{Methodology}
\label{sec:method}
To enable AI systems to interact effectively with domain experts' knowledge bases in astrophysics, we present a comprehensive framework consisting of four integrated components designed to systematically evaluate RAG Agents.

\subsection{CosmoPaperQA: Benchmark for Authentic Research Scenarios}
To address the evaluation challenges identified in the previous section, we manually construct CosmoPaperQA.

We systematically selected five highly influential papers spanning critical areas of modern cosmology: the Planck 2018 cosmological parameters \cite{2020Planck}, CAMELS machine learning simulations \cite{Villaescusa_Navarro_2021, Villaescusa_Navarro_2022}, local Hubble constant measurements \cite{Riess_2016}, and recent Atacama Cosmology Telescope constraints \cite{calabrese2025atacamacosmologytelescopedr6}. This curation ensures comprehensive coverage of observational, theoretical, and computational aspects of modern cosmological research.

A team of expert cosmologists generated 105 question-answer pairs through a rigorous protocol designed to mirror research inquiries. The questions in our dataset span multiple complexity levels: (1) factual retrieval requiring specific parameter extraction, (2) synthetic reasoning requiring integration across multiple evidence sources, and (3) analytical interpretation requiring deep domain knowledge. Each pair underwent expert validation to ensure scientific accuracy and representativeness of real research scenarios, distinguishing our benchmark from synthetic alternatives that lack authentic complexity.

Hence, CosmoPaperQA is designed for the following evaluations: \textbf{zero-shot learning,} answering without prior training on specific question types; \textbf{open-ended questions,} mirroring research scenarios; and \textbf{multi-source knowledge synthesis,} requiring integration across observational, theoretical, and computational domains.

\subsection{SciRag: RAG Implementation Pipeline}
Our preprocessing pipeline addresses the requirements of astronomical literature through multi-stage processing.  Optical character recognition (OCR) integration using Mistral's advanced capabilities \cite{mistral_ocr_2025} handles tables, figures, mathematical expressions, and specialized notations common in astrophysics papers. This system generates multiple output formats, ensuring compatibility across different RAG backends. 

Document segmentation employs LangChain with 5000-token chunks and 250-token overlap, optimized for scientific text coherence. Special handling accommodates the 4096-token constraint of OpenAI Assistant while maintaining consistency across all implementations.

All RAG systems perform retrieval over the complete corpus of 5 papers, regardless of which paper a specific question was derived from. This design tests the system's ability to identify and retrieve relevant information from the correct source paper among multiple cosmological documents.

We evaluate nine RAG implementations spanning commercial APIs (OpenAI, VertexAI), hybrid architectures (ChromaDB with OpenAI/Gemini embeddings), academic tools (PaperQA2), and search-enhanced systems (Perplexity). All systems use temperature=0.01 and top-k=20 for consistent evaluation. Detailed analysis is in Appendix \ref{app: config}.

\subsection{Dual Evaluation Framework: Human Expert and Calibrated AI Assessment}

To evaluate the quality of RAG Agents' responses in cosmological research contexts, we compare generated answers against expert-validated ground-truth responses to determine whether core factual claims in generated responses align with ground-truth.

While a single domain expert would be the optimal evaluator for this evaluation task, human-expert evaluation faces critical scalability limitations that make it impractical to evaluate across multiple RAG Agents. To address this scalability challenge, we implement a calibrated LLMaaJ system for automated response evaluation. However, we maintain scientific rigor by conducting parallel human expert evaluations on our benchmark results to validate the AI judges' performance and ensure assessment quality. Detailed evaluation setup is in Appendix \ref{app: eval}. After obtaining the scores, we scaled them to 0-100 for comparison between different system configurations.

\section{Results}
\label{sec:Results}

\subsection{Human Evaluated Results}

From the expert-evaluated results, we observe that the top-performing ones (OpenAIPDF, OpenAI, VertexAI) are all commercial RAGs, achieving 86.7-91.4\% accuracy. Both hybrid implementations (HybridOAIGem: 85.7\% , HybridGemGem: 84.8\% ) achieve performance competitive with commercial RAGs. PaperQA2 (81.90\%) demonstrates solid performance but lags by 4.8-9.5 \% compared to top performers. The poor performance of Perplexity Assistant (17.1\%) and Gemini Assistant (16.2\%) shows that unfiltered web search and non-RAG integration are insufficient for expert-level scientific inquiry, reinforcing the essential role of RAG Agents in scientific knowledge synthesis for autonomous scientific discovery. These clear performance distinctions between different system architectures validate CosmoPaperQA as an effective benchmark for distinguishing RAG agents' capabilities in authentic scientific research scenarios.

\subsection{AI Evaluated Results} 

\textbf{Evaluation Concordance:} Both OpenAI and Gemini judges preserve the performance ranking observed in human evaluation. The performance gaps are preserved: baseline systems achieve 11.4-18.1\% (OpenAI judge) and 16.2-31.4\% (Gemini judge), while top-performing agents reach 80.0-84.8\% (OpenAI judge) and 88.6-91.4\% (Gemini judge). 

\textbf{Judge-Specific Patterns:} The OpenAI judge demonstrates conservative scoring, consistently rating systems 2-8\% lower than human experts across all categories. In contrast, the Gemini judge exhibits systematic overrating, scoring systems 5-15 percentage points higher than human evaluation (e.g., Gemini Baseline: 27.6\% vs Human: 16.2\%, Modified PaperQA2: 81.9\% vs Human: 73.3\%). This overrating pattern suggests that Gemini judge may be overly optimistic in assessing scientific accuracy.

For researchers seeking robust performance estimates, the OpenAI judge's conservative scoring provides a safer lower bound for system capabilities, while Gemini's optimistic scoring may overestimate real-world performance. 
Despite these systematic biases, the consistent ranking order across all three evaluation methods (Pearson $r > 0.99$) demonstrates the robustness of our assessment framework. VertexAI demonstrates superior cost-efficiency while maintaining strong performance, while OpenAI achieves highest accuracy at a greater operational cost. Detailed cost analysis is provided in Appendix \ref{app:Cost}.

\section{Discussion and Future Work}
\label{sec:dis}
While CosmoPaperQA represents a first step in systematic astronomical RAG evaluation, several design choices warrant discussion. Many questions explicitly reference their source papers (e.g., \textit{Cosmology From One Galaxy?} questions mention the paper title, others reference \textit{Planck 2018} or \textit{ACT DR6}). This was intentionally adopted to ensure clear answer provenance and facilitate rigorous evaluation. However, researchers typically formulate queries around scientific concepts without specifying source documents, and our explicit references may systematically improve RAG performance by providing retrieval cues. 

Additionally, our five-paper corpus, while enabling expert evaluation, is more constrained than typical research contexts where systems must search thousands of papers or use web search, likely leading to degraded retrieval performance due to increased noise and irrelevant content. Future iterations should incorporate naturalistic question formulations and progressively larger document collections to test systems' ability to identify relevant sources without explicit guidance and understand how accuracy scales with corpus size.

Our results also reveal important insights into retrieval mechanisms that drive performance differences. OpenAI Assistants (89.5-91.4\%) use OpenAI's file search tool, which combines automatic query rewriting, parallel searches, keyword and semantic search, and result reranking. This multi-faceted approach outperforms simple semantic-only retrieval used in hybrid systems (84.8-85.7\%). Future work should evaluate domain-specific retrieval enhancements such as hybrid sparse-dense methods, contextual chunk expansion, query decomposition strategies, and multi-hop reasoning approaches to further optimize RAG performance for scientific applications.

The calibrated LLMaaJ evaluators developed in this work enable the next phase of our research: building AI questioner systems that can automatically generate domain-specific questions. Our current dataset of 945 human-evaluated responses provides a valuable training foundation for developing such automated question generation capabilities, potentially scaling evaluation to much larger document corpora.

The evaluation framework could be extended to other scientific domains such as chemistry, biology, or materials science to demonstrate generalizability. Despite these limitations, our framework provides a foundation for more comprehensive astronomical RAG benchmarks.

\section{Conclusion}
\label{sec:conclusion}

We have evaluated 9  agent configurations on 105 Cosmology Question-Answer (QA) pairs that were built specifically for this purpose, based on 5 carefully selected papers. The papers were selected for their impact on the field and the quality of the presentation of their results, and their relevance to the autonomous discovery systems that we are building, e.g., cmbagent, presented in a companion paper.  

The 9 agent configurations were manually evaluated by a human expert with more than 10 years of experience in the field, that is, a total of 945 generated answers were assessed. We find that currently the best RAG agent configuration uses OpenAI embedding and generative models, achieving 91.4\% accuracy. VertexAI (86.7\%) and hybrid architectures (84.8-85.7\%) demonstrate competitive performance. These configurations outperform academic tools uch as PaperQA2 (81.9\%), \cite{lála2023paperqaretrievalaugmentedgenerativeagent,skarlinski2024languageagentsachievesuperhuman}, which we attribute to the summarization steps in such systems that may dilute specific factual information critical for our evaluation tasks.  Notably, online tools like Perplexity perform poorly (17.1\%), showing essentially no advantage over frontier LLMs without RAG (16.2\%), indicating that unfiltered web search is insufficient for expert-level scientific inquiry.

Using our human evaluation results, we are able to calibrate evaluator agents which can be used as robust proxy for human evaluation. These results allow us to systematically select the best RAG agent configuration for multi-agent system for autonomous scientific discovery in astrophysics and provide us with AI evaluators that can be scaled to much larger evaluation datasets. By themselves, our 945 manually evaluated QA pairs constitute a precious dataset that can serve for the calibration of future AI evaluator agents. 



\section*{Impact Statement}
This paper presents work whose goal is to advance the field of Machine Learning. 
There are many potential societal consequences of our work, none which we feel must be 
specifically highlighted here.

\section*{Author Contributions}

XX led the work and wrote the paper. BB led the work, supervised XX and AD, and provided the human evaluation for all the 945 answers. AD created the CosmoPaperQA benchmark dataset. AL, FVN, LX and IZ provided crucial input at various stages of this work.

\section*{Acknowledgments}

The work of BB was partially funded by an unrestricted gift from Google, the Cambridge Centre for Data-Driven Discovery Accelerate Programme and the Infosys-Cambridge AI Centre.  We are very grateful to the referees and panel of the ICML 2025 ML4ASTRO workshop for reviewing and accepting our work.

\newpage
\bibliography{example_paper}
\bibliographystyle{icml2025}

\newpage
\appendix
\onecolumn
\section{Detailed System Configurations}
\label{app: config}
All generation agents are configured with a temperature of 0.01 for consistent, deterministic responses, and top-k=20 (retrieving the 20 most similar document chunks per query) excluding Gemini Assistant, PaperQA2 (both versions) and Perplexity Assistant. The implementation provides both semantic search and hybrid retrieval capabilities across different backends, with specific configurations optimized for each system's strengths. Here are the configurations that we use for each assistant.

\textbf{OpenAI Assistant:} Direct implementation of OpenAI vector stores with file search tool (providing automatic query rewriting, parallel searches, keyword+semantic search, and result reranking) with text-embedding-3-large \cite{openai2023embeddings} for embeddings and GPT-4.1 for generation, with configurable retrieval parameters (similarity threshold=0.5).

\textbf{OpenAIPDF Assistant:} Direct PDF processing implementation without OCR preprocessing, enabling comparison of raw PDF handling versus OCR-enhanced document processing. Identical configuration to OpenAI Assistant, but operates on unprocessed PDF documents.

\textbf{VertexAI Assistant:}  Google Cloud implementation using Google's text-embedding-005 for embeddings and gemini-2.5-flash-preview-05-20 \cite{google2023geminipro} for generation. Creates RAG corpora through Vertex AI infrastructure with automatic document ingestion from Google Cloud Storage buckets. Supports semantic search with configurable similarity thresholds (0.5).

\textbf{Gemini Assistant:}  Direct integration with Google's Gemini model gemini-2.5-flash-preview-05-20 for baseline comparison without specialized RAG infrastructure. 

\textbf{HybridGemGem Assistant:} Dual-Gemini implementation using Gemini's text-embedding-001 for embedding, leading embedding model on MTEB \cite{muennighoff-etal-2023-mteb} \footnote{Retrieved on 30-05-2025}with ChromaDB storage and gemini-2.5-flash-preview-05-20 for generation. Supports ChromaDB backends with semantic-only search.

\textbf{HybridOAIGem Assistant:} Cross-platform architecture identical to HybridGemGem but specifically configured with OpenAI embeddings (text-embedding-3-large) and gemini-2.5-flash-preview-05-20, enabling comparison of embedding-generation combinations.

\textbf{PaperQA2:} Standard academic RAG implementation utilizing GPT-4.1 across all components (search, summarization, retrieval), evidence retrieval k=30, maximum 5 citations per response (optimal settings from original work). Processes OCR-enhanced documents with semantic-only search.

\textbf{Modified PaperQA2:} Domain-adapted version with identical technical configuration but specialized astronomical prompts and cosmological citation protocols. Uses evidence retrieval k=10 (reduced from standard k=30) for more focused responses.

\textbf{Perplexity Assistant:} Web-search enabled system using sonar-reasoning-pro model with real-time access to current literature. No local vector storage - relies entirely on web retrieval.

This diverse implementation suite enables comprehensive comparison across commercial, academic, and hybrid approaches, providing empirical guidance for selecting optimal RAG configurations for autonomous scientific discovery workflows.
\section{Evaluation Setup}
\label{app: eval}
A domain expert is provided (1) a question query, (2) an ideal solution validated by experts, and (3) an RAG Agent-generated response. Then, evaluation is based on

\textbf{Correct (1):} Generated responses demonstrate factual accuracy, and capture essential scientific understanding equivalent to the ideal answer.

\textbf{Incorrect (0):}  Generated responses contain errors, contradict established scientific knowledge, or fail to include all the core concepts of ideal answers.

After obtaining the scores, we scaled them to 0-100 for comparison between different system configurations.

The cosmologist who evaluated the response is a domain expert with a PhD-level degree currently working as a researcher in astronomy, astrophysics, or physics. Together with this cosmologist, we designed the evaluation criteria and pipeline to ensure alignment with authentic research standards. In total,  our expert evaluated 945 responses (9 systems $\times$ 105 questions) generated by RAG Agents.

We explored LLM-as-a-Judge (LLMaaJ) \cite{gu2025surveyllmasajudge,zheng2023judgingllmasajudgemtbenchchatbot}, an AI-based evaluation system calibrated for scientific research queries, using a binary scoring protocol aligned with human expert methodology. Our prompting experiments in Appendix \ref{app: CoT Prompt} revealed that chain-of-thought, which asks models to formulate their underlying reasoning process, typically enhances evaluation accuracy and improves concordance with field expert judgments.

To investigate the bias of the pipeline specifically, as LLM evaluators may prefer responses generated by themselves \cite{bias_in_AI}, we used two LLM-as-a-Judge settings. Given that majority of generation systems utilize either OpenAI or Gemini-based agents, with the exception of the Perplexity Agent, we used the OpenAI o3 mini and Gemini gemini-2.5-pro-preview-06-05, reasoning models for evaluation.

\begin{figure} [htbp]
    \centering
    \includegraphics[width=1.0\linewidth]{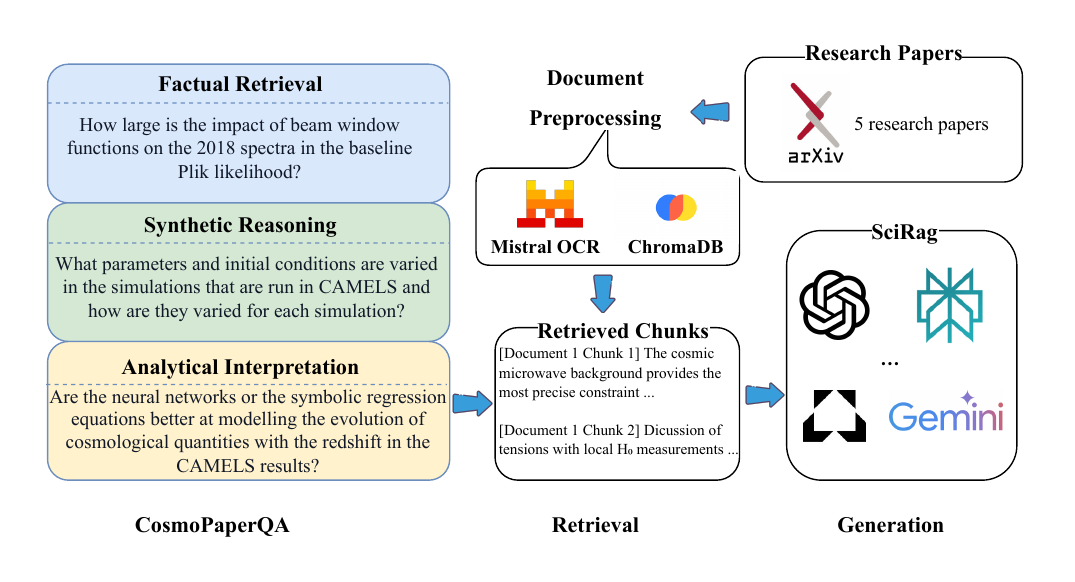}
    \caption{SciRag System Architecture and CosmoPaperQA Benchmark Overview. Our framework integrates document preprocessing, retrieval mechanisms, and multi-provider generation to enable systematic evaluation of RAG Agents on astronomical literature. }
    \label{fig:scirag}
\end{figure}

\section{RAG Prompts}
\label{app:rag_prompts}
Our modified PaperQA2 prompt priorities conciseness and domain specificity for efficient human evaluation.
\begin{tcolorbox}[colback=blue!5!white, colframe=blue!75!black, title={\textbf{Modified PaperQA2 Prompt}}]
Provide a concise answer in 1-2 sentences maximum. 

Context (with relevance scores):\{context\}

Question: \{question\}

Write a concise answer based on the context, focusing on astronomical facts and concepts. 
If the context provides insufficient information, reply

\{CANNOT\_ANSWER\_PHRASE\}.

Write in the style of a scientific astronomy reference, with precise and 
factual statements. The context comes from a variety of sources and is 
only a summary, so there may be inaccuracies or ambiguities. 

\{prior\_answer\_prompt\} Answer (maximum one sentence):
\end{tcolorbox}
\noindent 
In contrast, the original prompt emphasizes comprehensive information synthesis, mandatory citation and Wikipedia-style formatting.
\begin{tcolorbox}[colback=blue!5!white, colframe=blue!75!black, title={\textbf{PaperQA2 Prompt}}]

Answer the question below with the context.

\bigskip
Context (with relevance scores):\{context\}

Question: \{question\}

Write an answer based on the context. 

If the context provides insufficient information reply \{CANNOT\_ANSWER\_PHRASE\}

For each part of your answer, indicate which sources most support 
it via citation keys at the end of sentences, like \{example\_citation\}. 

Only cite from the context above and only use the citation keys from the context. 
\{CITATION\_KEY\_CONSTRAINTS\}

Do not concatenate citation keys, just use them as is. 

\bigskip
Write in the style of a Wikipedia article, with concise sentences and 
coherent paragraphs. The context comes from a variety of sources and is 
only a summary, so there may inaccuracies or ambiguities. If quotes are 
present and relevant, use them in the answer. This answer will go directly 
onto Wikipedia, so do not add any extraneous information.

\{prior\_answer\_prompt\}

Answer (\{answer\_length\}):
\end{tcolorbox}
The Hybrid SciRag assistant adopt a structured approach, requiring a JSON format return for consistent response parsing.
\begin{tcolorbox}[colback=blue!5!white, colframe=blue!75!black, title={\textbf{Hybrid Assistants Prompt}}]
You are a helpful assistant. Answer based on the provided context. 
You must respond in valid JSON format with the following structure:

\bigskip
\{
  "answer": "your detailed answer here",
  "sources": ["source1", "source2", "source3"]\}
\bigskip

The sources must be from the **Context** material provided. 
Include source names, page numbers, equation numbers, table 
numbers, section numbers when available. Ensure your response 
is valid JSON only.
\end{tcolorbox}
The Perplexity assistant uses web search to specific papers while utilizing its real-time retrieval capabilities.
\begin{tcolorbox}[colback=blue!5!white, colframe=blue!75!black, title={\textbf{Perplexity Assistants Prompt}}]
You are a scientific literature search agent specializing in cosmology.

We perform retrieval on the following set of papers:
\{paper\_list\}

Your task is to answer questions using ONLY information from these specific papers. 

Do not use any other sources or general knowledge beyond what these papers contain.

\bigskip
Instructions:

1. Search for information relevant to the question within the specified papers

2. Provide a CONCISE answer in EXACTLY 1-3 sentences. Do not exceed 3 sentences under any circumstances.

3. Add numerical references [1], [2], [3], etc. corresponding to the paper numbers listed above

4. If the papers don't contain sufficient information, state this clearly in 1-2 sentences maximum

5. Focus ONLY on the most important quantitative results or key findings

6. Be precise, direct, and avoid any unnecessary elaboration or context

\bigskip
CRITICAL: Your answer section must contain no more than 3 sentences total. Count your sentences carefully.

You must search your knowledge base calling your tool. The sources must be from the retrieval only.

Your response must be in JSON format with exactly these fields:

- "answer": Your 1-3 sentence response with citations

- "sources": Array of citation numbers used (e.g., ["1", "2"])

\end{tcolorbox}
 Gemini Assistant's approach to leveraging pre-trained knowledge of specific cosmological papers without requiring external retrieval mechanisms.

\begin{tcolorbox}[colback=blue!5!white, colframe=blue!75!black, title={\textbf{Gemini Assistant Prompt}}]
You are a scientific literature agent specializing in cosmology.

You have access to the following key cosmology papers in your knowledge base:
\{paper\_list\}

Your task is to answer cosmology questions using your knowledge of these papers and general cosmology knowledge.
Instructions:
1. Answer the question based on your knowledge of cosmology and the listed papers

2. Provide a CONCISE answer in EXACTLY 1-2 sentences maximum

3. Add numerical references [1], [2], [3], etc. when citing the specific papers listed above

4. Focus ONLY on the most important quantitative results or key findings

5. Be precise, direct, and avoid any unnecessary elaboration

\bigskip
Paper reference guide:

[1] - Planck 2018 cosmological parameters

[2] - CAMELS machine learning cosmology simulations  

[3] - Single galaxy cosmology analysis

[4] - Local Hubble constant measurement (Riess et al.)

[5] - Atacama Cosmology Telescope DR6 results

\bigskip
CRITICAL: Your answer must be no more than 2 sentences total. Count your sentences carefully.

Your response must be in JSON format with exactly these fields:

- "answer": Your 1-2 sentence response with citations

- "sources": Array of paper citations [1]-[5] that are relevant to your answer
\end{tcolorbox}

OpenAI/VertexAI assistants use a tool-based retrieval approach with markdown formatting, emphasising precise source and knowledge integration.
\begin{tcolorbox}[colback=blue!5!white, colframe=blue!75!black, title={\textbf{OpenAI/Vertex Assistants Prompt}}]
You are a retrieval agent. 
You must add precise source from where you got the answer.
Your answer should be in markdown format with the following 
structure: 

**Answer**:\{answer\}

**Sources**:\{sources\}

You must search your knowledge base calling your tool. The 
sources must be from the retrieval only.
You must report the source names in the sources field, if 
possible, the page number, equation number, table number, 
section number, etc.
\end{tcolorbox}

\section{CoT Prompts}
\label{app: CoT Prompt}
AI judges are given the following prompt:
\begin{tcolorbox}[colback=blue!5!white, colframe=blue!75!black, title={\textbf{Judge Prompt}}]

You are an expert scientific evaluator assessing the quality of scientific responses against reference answers.

\bigskip
Your task is to evaluate responses using one critical criterion:

ACCURACY (0-100):

CRITICAL: Use ONLY these two scores for accuracy:

- 100: The answer contains the core correct factual content, concepts, and conclusions from the ideal answer

- 0: The answer is fundamentally wrong or contradicts the ideal answer

This is a BINARY evaluation - either the answer is essentially correct (100) or fundamentally incorrect (0).

No partial credit or intermediate scores allowed.

\bigskip
EVALUATION GUIDELINES:

- Focus ONLY on whether the main scientific concepts and conclusions are correct

- Check that the core factual claims from the ideal answer are present in the generated answer

- Verify the overall conceptual direction and main conclusions align

- Additional correct information beyond the ideal answer is acceptable

- Only award 0 if the answer contradicts the ideal answer or gets the main concepts wrong

- Award 100 if the answer captures the essential correct scientific understanding

\bigskip
Provide your evaluation with the numerical score and detailed rationale explaining why you chose 100 or 0."""

\bigskip
Please evaluate this system's response against the ideal answer:

QUESTION: \{question\}

GENERATED ANSWER:

\{generated\_answer\}

IDEAL ANSWER:

\{ideal\_answer\}

\bigskip
Evaluate based on:

Accuracy (0-100): How factually correct is the answer compared to the ideal?

Use the evaluate\_response function to provide your structured evaluation with detailed rationale.

\end{tcolorbox}
\section{Cost Performance Analysis}
\label{app:Cost}
Cost considerations are critical for scientific research deployment, where institutions face budget constraints and researchers require sustainable access to AI-powered literature analysis tools. While our evaluation represents a controlled academic setting, understanding cost-performance trade-offs enables informed decisions for scaling RAG systems across research groups, institutions, and broader scientific communities.
\begin{table}[h]
\centering
\label{tab:cost_analysis}
\begin{tabular}{lcc}
\toprule
\textbf{Assistant Configuration} & \textbf{Cost per Query (\$)} & \textbf{Cost Ratio} \\
\midrule
VertexAI Assistant & 0.000357 & 1.0$\times$ \\
HybridOAIGem Assistant & 0.003182 & 8.9$\times$ \\
HybridGemGem Assistant & 0.003806 & 10.7$\times$ \\
Gemini Assistant (Baseline) & 0.004692 & 13.1$\times$ \\
Perplexity Assistant & 0.005200 & 14.6$\times$ \\
OpenAI Assistant & 0.048798 & 136.7$\times$ \\
OpenAIPDF Assistant & 0.048798 & 136.7$\times$ \\
PaperQA2 & 0.048798 & 136.7$\times$ \\
Modified PaperQA2 & 0.048798 & 136.7$\times$ \\
\bottomrule
\end{tabular}
\caption{Cost Analysis and Efficiency Comparison for RAG Configurations}
\end{table}

Table \ref{tab:cost_analysis} reveals substantial cost variations across configurations. VertexAI demonstrates exceptional cost efficiency (\$0.000357 per query) while maintaining strong performance (86.7\% accuracy), making it ideal for large-scale deployment. For our 105-question evaluation, total costs ranged from \$0.037 (VertexAI) to \$5.12 (GPT-4.1 based systems), representing a 137$\times$ cost difference.

The cost differences reflect underlying model pricing structures: GPT-4.1 costs \$0.002 per 1K input tokens and \$0.008 per 1K output tokens, while Gemini 2.5 Flash charges \$0.00015 per 1K input tokens and \$0.0006 per 1K output tokens. For a typical research corpus of 1,000 papers with 10,000 queries, projected costs would range from \$35.7 (VertexAI) to \$4,880 (OpenAI systems).

Hybrid approaches (HybridOAIGem: \$0.003182, HybridGemGem: \$0.003806) provide compelling cost-performance balance, achieving 84.8-85.7\% accuracy while reducing costs by 93\% compared to OpenAI systems. This positions them as practical solutions for resource-constrained research environments requiring both high accuracy and operational sustainability.

\begin{figure*}[!htbp]
    \centering
    \includegraphics[width=1.0\linewidth]{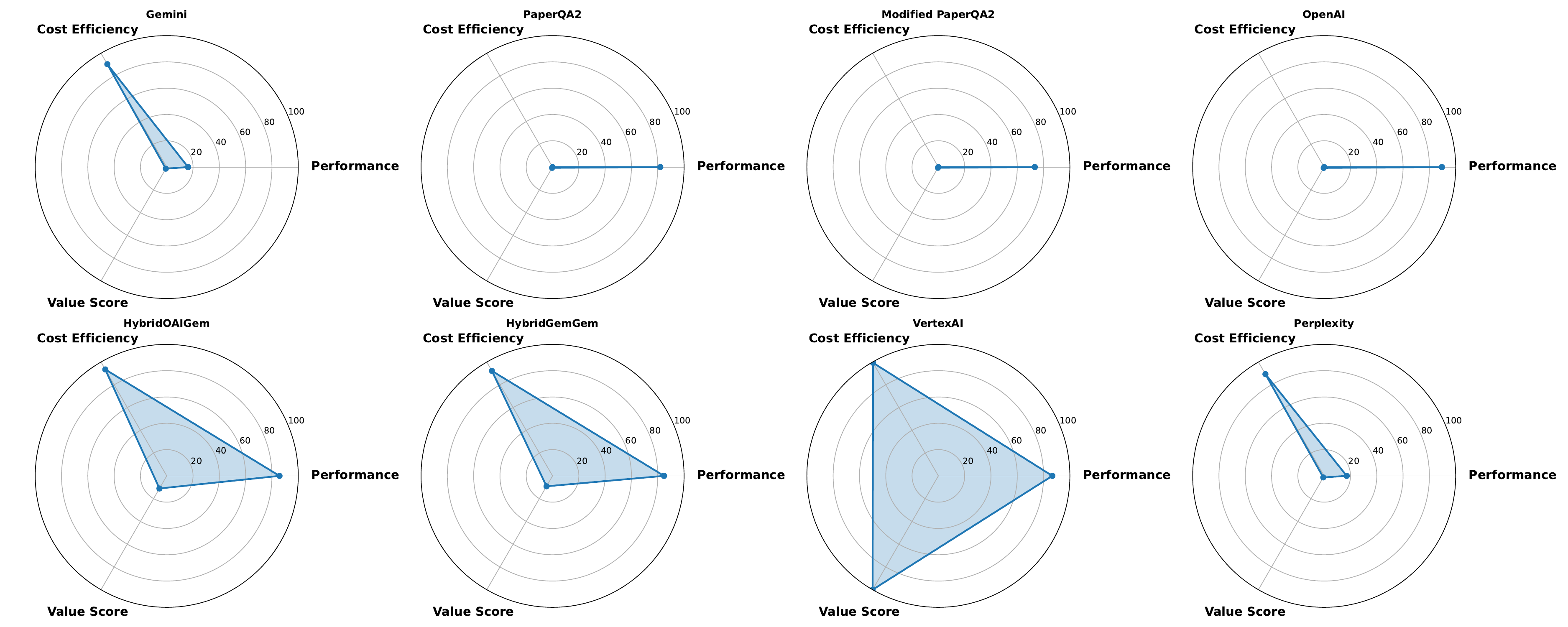}
    \caption{Multi-dimensional performance analysis of SciRag Agents across three key metrics: Performance (accuracy score), Cost Efficiency (inverse of operational cost), and Value Score (performance per unit cost). Each radar chart represents one agent, with larger areas indicating better overall value. Cost estimates are approximated using identical queries across different SciRag Agents for comparison.}
    \label{fig:cost-efficiency-analysis}
\end{figure*}

Figure \ref{fig:cost-efficiency-analysis} synthesizes these trade-offs across performance, cost efficiency, and overall value. While OpenAI systems achieve highest accuracy (89.5-91.4\%), their poor cost efficiency limits practical deployment scalability. Conversely, VertexAI maximizes value by combining strong performance with exceptional cost efficiency, making it suitable for widespread institutional adoption.

\end{document}